\documentclass[twocolumn]{aastex631}
\usepackage{tabularx}
\usepackage{amsmath}
\usepackage{amssymb}

\begin{document}

\title{Detecting the 21 cm Signal from the Cosmic Dark Ages}

\author{Willow Smith}
\affiliation{Department of Physics, Brown University, Providence, RI, USA}

\author{Jonathan C. Pober}
\affiliation{Department of Physics, Brown University, Providence, RI, USA}

\begin{abstract}

The cosmic ``Dark Ages'' is the period between the last scattering of the Cosmic Microwave Background (CMB) and the appearance of the first luminous sources, spanning redshifts $1100\gtrsim z\gtrsim 30$. The only way to observe this period is by examining the 21\,cm hyperfine transition line of neutral hydrogen H{\small I}, which --- given the high redshifts (and hence long wavelengths) --- must be observed from outside the Earth's ionosphere. Given the faintness of the signal, concepts for a radio array on the lunar far side (where large collecting areas can be deployed and radio frequency interference is minimal) have been proposed, like FarView or FARSIDE, but designs are still in the preliminary stages. This paper studies multiple aspects of array design to determine the impact of different design decisions on sensitivity to the Dark Ages 21\,cm power spectrum. We do so by using the sensitivity package \texttt{21cmSense} to model and simulate various array configurations. We present a fiducial design based on a modification of the FarView concept, which consists of a collecting area of $\sim2.5\,\rm{km}^{2}$ with 82,944 tightly packed dual-polarization dipoles grouped into 5,184 correlated elements, or subarrays, delivering a $>10\sigma$ detection of the $z=30$ signal with a five year lifetime. We find that, beyond mere collecting area, the most important factor in achieving this sensitivity is the presence of very short baselines that can only be realized with small, closely packed antennas.

\end{abstract}

\section{Introduction} \label{sec:intro}

The cosmic Dark Ages begins immediately following the recombination epoch which terminates in the surface of last scattering that gives way to the Cosmic Microwave Background (CMB) and lasts until the Cosmic Dawn when the first star and galaxy formations lead to the Epoch of Reionization (EoR) \citep{Pritchard_2008}. This roughly corresponds to the redshift range $1100\gtrsim z\gtrsim 30$ according to standard models of first star formation \citep{2002Sci...295...93A,Bromm_2002,Crosby_2013}. The period is currently unobserved and often regarded as a possible treasure trove of statistical measurements of the underlying cosmological parameters of the universe.

The lack of luminous sources makes the Dark Ages extraordinarily difficult to study through direct observation. The only available probe of this era is the 21\,cm hyperfine transition line of neutral Hydrogen (H{\small I}) \citep{Pritchard_2012}. This emission, when observed from the current day, is redshifted well into the low-frequency radio range ($\sim10-40$\,MHz). This range of redshifted 21\,cm emission presents both useful and challenging aspects of observation. Helpfully, the large frequency range allows us to study a 3D tomography of the 21\,cm line throughout the Dark Ages which in turn offers a greater number of accessible modes of cosmological Fourier space than the CMB \citep{Loeb_2004}. 

Unhelpfully, the redshifted Dark Ages signal is primarily at frequencies where the Earth's ionosphere is opaque \citep{Vedantham_2016}. Coupled with a substantial overlap with artificial radio frequency interference (RFI), this aspect of the Dark Ages signal strongly necessitates a space-based radio instrument to observe it. The lunar far side is an ideal location for its ability to shield from Earth-based RFI and for its lack of an appreciable atmosphere, which allows any future instrument to access all Dark Ages Frequencies \citep{burns2021global21cmcosmologyfarside}.

The types of radio instruments that have been proposed for the moon include both single dish antennas --- like the Lunar Crater Radio Telescope (LCRT) \citep{9438165} --- and radio interferometric arrays. However, radio interferometric arrays like FARSIDE \citep{burns2019farsidelowradiofrequency}, FarView \citep{polidan2024farviewinsitumanufacturedlunar}, and CoDEX \citep{koopmans2019peeringdarkageslowfrequency} are the more common instruments found in the literature. These concepts range from small pathfinder arrays with $\mathcal{O}(100)$ antennas to arrays equal to and exceeding the Square Kilometre Array (SKA) in size with $\mathcal{O}(10^{5})$ antennas in a dense core. The exact layout of antennas also differs between designs with demands for appropriate spacing for lunar rover operation and for connection and distribution of power through solar batteries. Finally, these designs tend to favor dipole antennas which are simple to construct and effective at low-frequency.

The primary probe of a Dark Ages radio interferometer experiment is the 21\,cm redshifted frequencies that map to a 21\,cm brightness temperature relative to the CMB, $T_{b}$, on the sky. $T_b$ is set by the neutral fraction of H{\small I} --- which is 1 throughout the Dark Ages --- the spin temperature $T_s$ relative to the CMB luminous background temperature $T_{\gamma}$, an optical depth dependent on the line-of-sight peculiar velocity, and most importantly for our measurements, the local density of H{\small I} $\delta_H$ \citep{Loeb_2004}. $\delta_H$ is best described in cosmological Fourier space by the power spectrum, briefly described analytically in Section \ref{subsec:power}.

The 21\,cm power spectrum is a powerful probe of density fluctuations in H{\small I} resulting from primordial perturbations from inflation. As mentioned previously, the available Fourier modes are far greater than the CMB's available modes which allows for the further constraining of cosmological parameters in the cosmic variance limit --- i.e. in the limit of negligible thermal noise \citep{Mondal_2023}. Additionally, a lack of astrophysical processes enables measurements and constraints on exotic physics like charged Dark Matter \citep{Mu_oz_2018}.

Therefore, in this paper, we seek to analyze a lunar far side array's sensitivity to the power spectrum. Although exotic physics models can significantly boost the power spectrum amplitude \citep{Mu_oz_2018}, we focus on the ``worst'' case scenario for a detection, i.e., standard physics and a correspondingly faint 21\,cm signal. In Section \ref{sec:concepts}, we go over some of the currently proposed designs and discuss the key parameters of array design that impact sensitivity. In Section \ref{sec:calc}, we briefly delve into the theory behind sensitivity to the Dark Ages 21\,cm power spectrum and discuss how the \texttt{21cmSense} sensitivity package is implemented. In Section \ref{sec:sim}, we present the bulk of our work in a step-by-step comparison of varying design parameters to a fiducial array analogous to those discussed in Section \ref{sec:concepts} and examine how the significance of detection of the power spectrum evolves. Finally in Section \ref{sec:conc}, we discuss caveats of the designs and future simulations that are necessary for a lunar far side radio interferometer. Unless otherwise stated, we assume a Planck 2018 cosmological model with $h=0.6766$ and $\Omega_{m}=0.30966$ as implemented in \texttt{astropy}'s cosmology realization \citep{2020A&A...641A...6P, astropy:2013,astropy:2018,astropy:2022}. 

\section{Lunar Far Side Array Concepts} \label{sec:concepts}

Interest in a lunar radio array goes back several decades with preliminary concepts being proposed as few as three decades ago \citep{Burns_1991, Lazio:2007zp}. Since then the 2013 NASA astrophysics roadmap \citep{kouveliotou2014enduringquestsdaringvisionsnasa} stressed the importance of a lunar far side radio telescope for mapping the cosmic dawn driving the discussion on a number of newer concepts. Recent instruments like ROLSES \citep{burns2021lowradiofrequencyobservations} have demonstrated the viability of radio astronomy from the moon, while future experiments like LuSEE will continue to push towards lunar far side measurements. \citep{bale2023luseenightlunarsurface}.

As mentioned previously, there are single dish radio telescopes like the LCRT \citep{9438165}, but we follow the direction set by ground-based experiments that study the Cosmic Dawn --- like HERA, MWA, and SKA \citep{Berkhout_2024,Wayth_2018,Weltman_2020} --- to use a large-scale radio interferometer to study the Dark Ages signal. 

We look to a handful of radio interferometer concepts for reference. FARSIDE presents a pathfinder array with 128 dual-polarization 20\,m dipoles in an asymmetric petal-like configuration for a total effective collecting area of 6700\,$\rm m^{2}$ at 10\,MHz \citep{burns2019farsidelowradiofrequency}, while CoDEX presents a generic densely packed core of antennas with a 5-year mission lifetime, 10\,MHz bandwidth, full-sky imaging, and three stages of collecting area: 1\,$\rm{km}^{2}$, 10\,$\rm{km}^{2}$, and 100\,$\rm{km}^{2}$ \citep{koopmans2019peeringdarkageslowfrequency}. Both array concepts offer useful designs for probing the Cosmic Dawn and Dark Ages signals in incremental stages, but it is the purview of this paper to study a more generic design along the lines of CoDEX. FarView offers a similar architecture to CoDEX with a more specific layout and location that makes it a useful candidate for sensitivity calculations \citep{polidan2024farviewinsitumanufacturedlunar}. In the next section (\ref{subsec:farview}), we will review the parameters of the FarView concept presented in \citet{polidan2024farviewinsitumanufacturedlunar}; in Section \ref{subsec:fiducial} we will in turn present a fiducial array for detections of the 21\,cm dark ages signal adapted from the FarView design.  

\subsection{FarView} \label{subsec:farview}

FarView is an in-situ manufactured radio interferometer intended to be on the order of 100,000 dipole antennas (hereafter, simply ``dipoles'') with roughly half of the antennas in a 36 $\rm{km}^2$ core and the other half as outriggers extending to 200 $\rm{km}^2$. We base our fiducial array, discussed in Section \ref{subsec:fiducial}, around the initial design concept of FarView so it is helpful to layout the key details of its design.

The primary design aspects of FarView we reference are the \emph{subarray} and the \emph{cluster}. Each of these concepts is part of a hierarchical structure based around engineering constraints. The \emph{subarray} is a collection of dipoles that, through the process of beamforming, are treated as a single antenna. In the original FarView concept --- which we refer to as the Advances in Space Research (ASR) concept laid out in \cite{polidan2024farviewinsitumanufacturedlunar} --- subarrays are anywhere from 300 to 600 dipoles in size. The implementation of subarrays is to circumvent the computationally difficult, if not impossible, task of correlating individual pairs of antennas which for an array of this size would be on the order of $N^2 \sim 10^{10}$ correlations.

In the ASR Concept, a subarray of 400 dipoles is powered by a single solar cell. In our study, we explore the possibility of reducing the subarray size, letting multiple subarrays be powered by one solar cell.  Hence, following internal conversations with the FarView team, we introduce the concept of a \emph{cluster}, distinct from a subarray. The cluster is defined as the number of dipoles that can be powered by a single solar cell and is a collection of subarrays. Unlike the subarray design, the cluster design does not originate from computational constraints and plays a limited role in the sensitivity calculations presented here. We choose not to omit the cluster concept altogether, however, as it helps standardize the geometric layout of the arrays simulated in Section \ref{sec:sim} and reflects the engineering constraints that ultimately must be part of any lunar array design.

A secondary design aspect of clusters we adapt into our analysis is the service road. A service road allows a rover to navigate the array for maintenance and deployment of dipoles. We therefore also include service roads in our arrays for applicability to practical designs.

The concepts of the subarray, cluster, and service road should not necessarily be unique to FarView. These designs aspects are purely to address limitations introduced by the massive size of an array necessary to observe the Dark Ages signal. Additionally, as discussed in Section \ref{subsec:21cmsense}, \texttt{21cmSense} abstracts away from these concepts, meaning our sensitivity calculations are applicable to any FarView-scale concept. Therefore, using FarView as a basis for our Fiducial array does not lose generality.

\begin{table}[htbp]
    \centering
    \begin{tabular}{|c|c|c|}
    \hline
        & Fiducial Array & ASR Concept\\
    \hline
        Total Dipoles in Core & 82,944 & 48,400 \\
    \hline
        Total Dipoles in Subarray & 16 & 400  \\
    \hline
        Dipole Spacing & 11.5\,m & 17\,m \\
    \hline
        Subarray Spacing & 11.5\,m & 500\,m \\
    \hline
    \end{tabular}
    \caption{Key differences between our fiducial array, described in Section \ref{subsec:fiducial}, and the Advances in Space Research (ASR) FarView paper concept. The increase in total dipoles, decrease in subarray size, and decrease in spacing all yield greater sensitivity for our fiducial array. Notably, clusters, and thus total dipoles in a cluster and service road width, are excluded from this table since the ASR concept has no need for them; a cluster and subarray are equivalent for the ASR concept.}
    \label{ASR}
\end{table}

\section{Calculating Sensitivities} \label{sec:calc}

\subsection{The 21\,cm Power Spectrum} \label{subsec:power}

In order, to calculate sensitivity of a radio instrument to the Dark Ages signal we first briefly define and describe how we determine the 21\,cm power spectrum. For a more in-depth review of the power spectrum and 21\,cm cosmology refer to \cite{Furlanetto_2006, Morales_2010, Liu_2020}.

Because the cosmic matter distribution is statistically very nearly homogeneous, isotropic, and Gaussian --- as pointed towards by observational constraints of the CMB \citep{refId0} ---
we can describe, to good approximation, the local density of H{\small I} via the 2-point function

\begin{equation}
    \langle \delta_H({\bf{k}}) \delta_H({\bf{k'}}) \rangle
    = (2\pi)^3 \delta ^3({\bf{k}} + {\bf{k'}})P(k)
    \label{twopoint}
\end{equation}

\noindent where $P(k)$ is the one-dimensional matter power spectrum in cosmological Fourier space as a function of wavenumber $k$. Throughout the paper, when we refer to sensitivity to the power spectrum, we are referring to the sensitivity to the dimensionless power spectrum $\Delta^{2}(k)=(k^{3}P(k))/(2\pi^{2})$\footnote{The 21\,cm dimensionless power spectrum actually has units of mK$^{2}$.}.

We generate a brightness temperature power spectrum at redshift $z=30$ in \texttt{21cmFast} \citep{10.1111/j.1365-2966.2010.17731.x,Murray2020}\footnote{\url{https://github.com/21cmfast/21cmFAST}} using a 2\,Gpc box with a 4\,Mpc voxel size. We enable spin temperature $T_{s}$ fluctuations but otherwise use the default parameters of 21cmFast v3.3.1.Notably, the lack of astrophysics during the dark ages makes the brightness temperature power spectrum a direct probe of the dark matter density. Therefore, we also generate the dark matter density power spectrum to confirm this notion, and after normalizing the density and brightness temperature power spectra to unity, we find the average percent difference between the curves to be 3.5\%. This indicates that at $z=30$ the density field is the principal driver of the 21\,cm brightness temperature and effects like ionization, heating, and velocities play a subdominant role. Therefore, we use the brightness temperature power spectrum in all plots.

\subsection{The 21\,cm Noise Power Spectrum} \label{subsec:theory}

The noise contribution to the 21\,cm power spectrum for a radio interferometric array was first derived in rigorous detail in \cite{Morales_2005} and \cite{McQuinn_2006}. We refer to this as the sensitivity throughout the paper i.e. lower noise means we are more sensitive to the 21cm Power Spectrum of the Dark Ages. 

There are two contributions to the noise power spectrum: thermal noise which we focus on in this section, and sample variance (typically referred to as cosmic variance). At Dark Ages frequencies, the sky brightness temperature is the dominant contributor to the thermal noise \citep{Mozdzen_2016}. On the other hand, the contribution from the Dark Ages power spectrum to sample variance is less significant. For example, in our fiducial array described in Section \ref{subsec:fiducial}, the sample variance is, at worst, an order of magnitude lower than thermal noise and on average 4 orders of magnitude lower. Therefore, we ignore sample variance, which is sub-dominant, in this paper.

In order to understand how various design parameters affect the sensitivity, we break down the analytical form of the thermal contribution to the noise power spectrum. For white-noise thermal fluctuations with root mean square (rms) brightness temperature, as shown in \cite{Parsons_2012}, the thermal noise contribution $\delta\Delta^{2}(k)$ to the dimensionless power spectrum $\Delta^{2}(k)$ can be written as

\begin{equation}
    \delta\Delta^{2}(k) \approx X^{2}Y \frac{k^{3}}{2\pi^{2}} \Omega B T^{2}_{\rm rms}
    \label{eq:Noise}
\end{equation}

\noindent Here $X^{2}Y$ is the cosmological scaling factor between observing coordinates or comoving volume, $\Omega$ is the primary beam field of view (FoV)\footnote{There is some nuance to what $\Omega$ refers to; in this case, we, and \texttt{21cmSense}, follow the formalism in Appendix B of \cite{Parsons_2014} to use $\Omega=\Omega_{p}^{2}/\Omega_{pp}$ for power spectrum sensitivity equations where $\Omega_{p}$ is the integral of the primary beam power and $\Omega_{pp}$ is is the integral of the square of the primary beam power.}, $B$ is the observing bandwidth, $k$ is the Fourier mode of the one-dimensional 21\,cm power spectrum, and $T_{rms}$ is the effective radiometric noise temperature for one $k$ mode. It is this form of the sensitivity equation that is used in \texttt{21cmSense}. 

We examine $T_{\rm rms}$ in Section \ref{subsec:21cmsense} in detail. The factors in front of $T_{\rm rms}$ in Equation \ref{eq:Noise} amount to a normalization constant to the power spectrum. $X^{2}Y$ is set by the redshift -- where we use $z=30$ throughout the paper -- and FLRW cosmological parameters. The field of view $\Omega=\lambda^{2} / A_{\rm eff}$ is set by the FoV of the smallest beam-formed receiver element, in this case a subarray with area $A_{\rm eff}$. 

\begin{figure*}[ht!]
\plotone{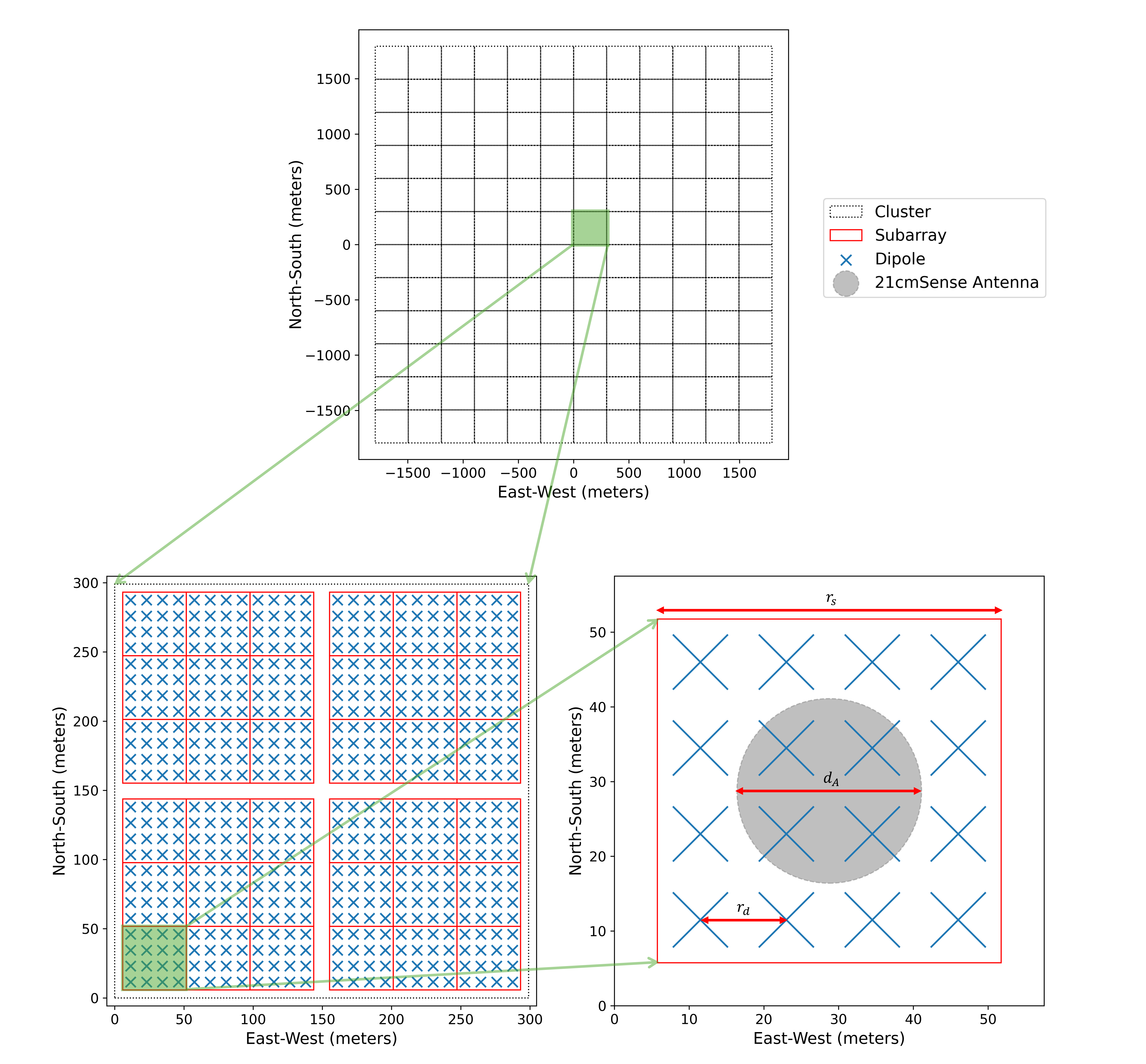}
\caption{The fiducial array. The top panel shows the entire array with each dotted black square representing a single cluster. The bottom-left panel shows the layout of one cluster as 4 groupings of subarrays (outlined in red) each separated by service roads with width equivalent to dipole spacing $r_{d}$. The bottom-right panel shows a single subarray of 16 x-shaped dipoles spaced center-to-center by $r_{d}$ with a total subarray width of $r_{s}$ and the abstracted \texttt{21cmSense} circular Antenna with diameter $d_{A}$ and equivalent $A_{\rm eff}$ to 16 dipoles. For the fiducial array $r_{d}=11.5\,\rm{m}$, $r_{s}=46\,\rm{m}$, and $d_{A}=24.7\,\rm{m}$.} 
\label{fig:fiducialArray}
\end{figure*}

\subsection{Array Modeling in 21cmSense} \label{subsec:21cmsense}

All sensitivities in this paper are calculated using version 2.0 of the python package \texttt{21cmSense} \citep{Murray2024}\footnote{\url{https://github.com/rasg-affiliates/21cmSense}}. \texttt{21cmSense} is based on the algorithm laid out in \cite{Pober_2013} and \cite{Pober_2014} and later summarized in \cite{Liu_2020}. 

In this section we are concerned with how \texttt{21cmSense} allows us to model a lunar farside array. Typically, \texttt{21cmSense} allows one to pass an array of antenna positions in $ENU$ (East-North-Up) coordinates in units of meters relative to a central location given by latitude. These antenna positions are then used to calculate baselines in units of wavelengths of the target frequency which then are used to construct a grid in $uv$ coordinates (the Fourier dual of the sky coordinates).

Importantly, we modify \texttt{21cmSense} to accept arrays at lunar coordinates. This involves first updating the interferometer python package dependency \texttt{pyuvdata} \citep{Hazelton2017} with the correct coordinate frames from the \texttt{lunarsky} package\footnote{\url{https://github.com/aelanman/lunarsky}} which functions as an extension to \texttt{astropy}. These coordinate frames are: the selenodetic frame for locations on the lunar surface which parallels the geodetic (longtiude, latitude, and altitude) frame of Earth; the Moon-centered, Moon-fixed (MCMF) frame for the global sky which parallels the Earth-centered, Earth-fixed (ECEF) frame also known as the geocentric (Cartesian X, Y, and Z) frame; and the lunar topocentric frame for the local sky which parallels the horizontal coordinate system (altitude and azimuth) on Earth. Integration with astropy allows us to convert the lunar topocentric and MCMF frames to the "space fixed" ICRS (right ascension and declination) frame with the appropriate rotational transformations. 

\texttt{Lunarsky} also defines the local sidereal time (LST) on the moon as the right ascension of local zenith in the ICRS frame at a given time, accounting for the moon's orbit about the Earth as the Earth orbits the sun. This coordinate functionality is then implemented in \texttt{21cmSense} \footnote{\url{https://github.com/rasg-affiliates/21cmSense/tree/adding-lunar-coords}}.  

For sensitivity to the power spectrum, we are concerned with the form of the rms brightness temperature:

\begin{equation}
    T_{\rm rms} = \frac{T_{\rm sys}}{\sqrt{2Bt}}
    \label{eq:Trms}
\end{equation}

\noindent where $T_{\rm sys}=T_{\rm sky} + T_{\rm rcv}$ is the contribution of the sky brightness temperature and receiver temperature, $t$ is the total integration time over $uv$ bins for a particular LST bin, $B$ is once again the bandwidth, and the factor of 2 in the square root accounts for both linear polarizations \citep{Parsons_2012}.

Just as on Earth, baselines on the moon sweep out elliptical tracks across the $uv$-plane --- only here, the timescale is a lunar day ($\sim 28$ Earth days) as opposed to 24 hours. These tracks allow \texttt{21cmSense} to perform rotation synthesis over a particular LST bin in the $uv$-plane. Importantly, the construction of the $uv$-plane only contributes to $t$ in $T_{\rm rms}$.  Therefore, the total integration time $t$ in Equation \ref{eq:Trms} is the primary contributing factor to sensitivity that any array's physical spacing and layout affects.

Because the moon is similarly behaved to the earth, we can enable tracking. Tracking is achieved when antennas center the primary beam on a particular RA of the sky and follow it through the night. This can be done by pointing the maximal response of an antenna at a desired source with steerable antennas or through beamforming in the case of fixed dipoles. Therefore, tracking offers a consistent and longer observation of a single field. Thus, by tracking a single RA through the night thermal noise can be substantially reduced. The alternative is to preform a drift-scan where the array is fixed on local zenith and allows multiple fields to "drift" through the primary beam. Therefore, by observing multiple fields drift-scan is more effective in reducing sample variance.

In \texttt{21cmSense}, antennas are assumed to be circular dishes. It is important to note that the effective area $A_{\rm eff}$ of an antenna is not necessarily equivalent to the physical area of the antenna but very nearly is for circular dishes with a Gaussian beam. It is, therefore, straightforward to convert a dipole of some specific $A_{\rm eff}$ to a circular dish with a radius $r$ corresponding to its physical area $\pi r^2=A=A_{\rm eff}$. We can extend this approximation to the subarray, treating a collection of $n$ dipoles with total effective area $nA_{\rm eff}$ as a circular antenna with $A=nA_{\rm eff}$. Therefore, we implement a Gaussian primary beam in \texttt{21cmSense} for all sensitivity calculations throughout the paper. This approximation is only valid in the case of sensitivity calculations where we are only concerned with $A_{\rm eff}$, and this approximation would not apply to more complex simulations.

\begin{table*}[htbp]
    \centering
    \begin{tabularx}{\textwidth}{|l|p{1.5cm}|p{1.5cm}|p{1.5cm}|p{1.1cm}|p{1.5cm}|p{1.1cm}|l|l|p{1.5cm}|}
    \hline
        Comparison & Array Dimensions & Subarray Dimensions & Subarrays/ Cluster & Total Dipoles & Collecting Area & Dipole Spacing & Tracking & Foregrounds & Significance of Detection  \\
    \hline
        Fiducial & 12x12 & 4x4 & 36 & 82,944 & 2.49\,km$^2$ & 11.5\,m & Yes & None & 10$\sigma$ \\
         (\S\ref{subsec:fiducial}) \& & {} & {} & {} & {} & {} & {} & No & None & 3.4$\sigma$ \\
        (\S\ref{subsec:fore}) & {} & {} & {} & {} & {} & {} & Yes & Subtraction & 4.9$\sigma$ \\
        {} & {} & {} & {} & {} & {} & {} & Yes & Avoidance & 0.41$\sigma$ \\
    \hline
        Subarray & 12x12 & 3x3 & 64 & 82,944 & 2.49\,km$^2$ & 11.5\,m & Yes & None & 14$\sigma$ \\
        Size(\S\ref{subsec:subarray}) & 12x12 & 6x6 & 16 & 82,944 & 2.49\,km$^2$ & 11.5\,m & Yes & None & 7.0$\sigma$ \\
        ($\sqrt{n}$ x $\sqrt{n}$) & 12x12 & 12x12 & 4 & 82,944 & 2.49\,km$^2$ & 11.5\,m & Yes & None & 3.5$\sigma$ \\
    \hline
        Dipole & 12x12 & 4x4 & 36 & 82,944 & 2.49\,km$^2$ & 17.0\,m & Yes & None & 5.4$\sigma$ \\
        Spacing & 12x12 & 4x4 & 36 & 82,944 & 2.49\,km$^2$ & 9.0\,m & Yes & None & 16$\sigma$ \\
         (\S\ref{subsec:spacing}) ($r_{d}$) & 12x12 & 4x4 & 36 & 82,944 & 2.49\,km$^2$ & 7.0\,m & Yes & None & 23$\sigma$ \\
    \hline
        Collecting & 10x10 & 4x4 & 36 & 57,600 & 1.73\,km$^2$ & 11.5\,m & Yes & None & 6.8$\sigma$ \\
        Area (\S\ref{subsec:area}) & 14x14 & 4x4 & 36 & 112,896 & 3.39\,km$^2$ & 11.5\,m & Yes & None & 15$\sigma$ \\
        ($A_{\rm{coll}}$) & 16x16 & 4x4 & 36 & 147,456 &4.42\,km$^2$ & 11.5\,m & Yes & None & 20$\sigma$ \\
        {} & 18x18 & 4x4 & 36 & 186,624 & 5.60\,km$^2$ & 11.5\,m & Yes & None & 26$\sigma$ \\
    \hline
    \end{tabularx}
    \caption{A table showing the fiducial array and its permutations discussed in Section \ref{sec:sim}, broken down by parameter comparison. Array dimensions are in clusters and subarray dimensions are in dipoles. Array dimensions with subarray dimensions and subarrays per cluster gives the total number of dipoles in the array which is also listed here. Dipole Spacing is the spacing between dipoles in each subarray as well as the width of service roads. We also list whether the array is tracking a fixed point and whether foreground models (described in Section \ref{subsec:fore}) are considered. Finally, the significance of detection to the $z=30$ power spectrum in $\sigma$ is given for each configuration. Note, the ASR concept differs significantly in its spatial configuration from the Fiducial array such that it is not readily compared with the same set of parameters and is omitted from this table.}
    \label{Arrays}
\end{table*}

\subsection{Signifcance of Detection} \label{subsec:detection}

Finally, we input a theoretical Dark Ages power spectrum into \texttt{21cmSense}, as described in Section \ref{subsec:power}, which allows us to calculate the significance of detection. This quantity is a Pythagorean sum of the signal-to-noise ratio (SNR) $\Delta^{2}(k)/\delta\Delta^{2}(k)$ over all Fourier modes in number of standard deviations $\sigma$. We use this quantity to compare the effectiveness of various array configurations in Section \ref{sec:sim}.

\section{Simulated Sensitivities} \label{sec:sim}

Building an array sensitive to the Cosmic Dark Ages 21\,cm signal involves several adjustable parameters. Typically the key parameters are the total collecting area of the array and receiving element size. As explained in Section \ref{subsec:farview}, we adopt subarrays as our smallest receiving element as opposed to individual dipole antennas. Therefore, receiving element size is determined by dipole size, which we keep fixed, and dipole number $n$. Additionally, while not directly contributing to receiving element size or collecting area, dipole spacing affects how tightly we can pack subarrays. Thus, in addition to total collecting area, we have the parameters of subarray size, denoted $\sqrt{n}$ x $\sqrt{n}$, and dipole spacing $r_{d}$. In Section \ref{subsec:fiducial}, we discuss our fiducial array, and in subsequent sections, we discuss how varying each of these parameters affects sensitivity: first with subarray size in Section \ref{subsec:subarray}, then dipole spacing in Section \ref{subsec:spacing}, and finally collecting area in Section \ref{subsec:area}. We also briefly touch on how foregrounds impact the sensitivity in Section \ref{subsec:fore}. Table \ref{Arrays} contains a complete list of array configurations described in this section and their significances of detection.

\subsection{The Fiducial Array} \label{subsec:fiducial}

Initially, the ASR concept of FarView was an ideal candidate for our fiducial array. Preliminary sensitivity forecasts look promising with an estimated $\sim 10\sigma$ detection for a $\sim1 \rm{km}^2$ array observing over 5 years --- at half-duty cycle because the radio bright sun is in the sky for half the time. However, these forecasts do not use rigorous numerical simulations, instead using an approximation of the noise under the assumption of a constant density of visibilities in the $uv$-plane \citep{koopmans2019peeringdarkageslowfrequency}. Thus, inputting this array into \texttt{21cmSense} offers a more clear picture of the ASR concept's sensitivity. Surprisingly, the initial forecasts are overly optimistic with \texttt{21cmSense} yielding a $0.12\sigma$ detection for the ASR concept --- the reasons for this result being so different from the initial prediction are discussed in Section \ref{sec:conc}.

Therefore, to obtain a substantial $>10\sigma$ detection, we modify the ASR concept in a number of ways to create our fiducial array --- these differences are highlighted in Table \ref{ASR}. Notably, we ignore the outriggers in the ASR Concept in this comparison and do not adapt them into our fiducial array. Outriggers are antennas, or subarrays, randomly distributed outside the core of the array and are primarily used for calibration, so they would have a minor contribution to sensitivity calculations if included. 

In contrast to the ASR concept, the number of dipoles in a subarray in our fiducial array are reduced by an order of magnitude down to 16 dipoles. This reduction increases the number of shorter baselines allowing us to have higher sensitivity than the ASR concept with a computationally feasible $N^2 \sim 10^{7}$ correlations. With the reduced subarray size, we then introduce the cluster concept, described in Section \ref{subsec:farview}, as a collection of subarrays totaling 576 dipoles. While a cluster is not necessarily required to contain exactly 576 dipoles (the particular constraints depend on the solar cell design), we maintain this number throughout the paper for ease of reference.

We show the layout of our fiducial array in Figure \ref{fig:fiducialArray}. We place the array at a -44.1 lunar latitude corresponding with the Pauli crater, one of the proposed locations for FarView. This location is sufficiently flat over 10\,km scales for the efficient manufacture of a large array \citep{polidan2024farviewinsitumanufacturedlunar}. Therefore our placement of clusters, denoted by dotted black squares, side-by-side in a 12x12 square grid $\sim 3.5$\,km in diameter is reasonable within instrument limitations. In Section \ref{subsec:area}, we vary the total collecting area $A_{\rm coll}$ by modifying the dimensions of the array in number of clusters to a maximum square grid diameter of $\sim 6.3$\,km, well within the flat crater area.

Each cluster is composed of 36 subarrays, outlined in red, linked to a solar cell (not depicted in the figure). Subsequently, each subarray is composed of $n$=16 evenly spaced dipoles, marked by blue X's, with a total side-length $r_{s}=46$\,m.  Technically, each unit labeled ``dipole" is a pair of crossed 10\,m dipoles to account for dual-polarization, with an $A_{\rm eff}=30\,\rm{m}^{2}$ per polarization. This differs from the ASR concept where dipoles are single-polarization and offers a factor of 2 increase in sensitivity. We place dipoles $r_{d}=11.5$\,m apart center-to-center and leave equivalent space for service roads, shown as the gaps between the four quadrants of subarrays in a cluster. Note the tip-to-tip spacing is shorter than $r_{d}$; in this case, it is $\sim 4.4$\,m.  This is relevant in our analysis in Section \ref{subsec:spacing} when we vary $r_{d}$. 

We also show what an antenna as implemented in \texttt{21cmSense} would look like in this geometry, represented by a grey transparent circle. The diameter $d_{A}=2\sqrt{A/\pi}=24.7$\,m is set by the antenna area $A=nA_{\rm eff}$ as described in Section \ref{subsec:21cmsense} where $n=16$ dipoles.

Beyond the physical layout of the array, there are a number of parameters that affect sensitivity. We vary two of these, namely tracking and foregrounds, for the fiducial array only. Because we are dominated by thermal noise, we simply note the significance of detection without tracking at the end of this section. In Section \ref{subsec:fore}, we discuss in more depth the effects of foregrounds on the fiducial array. 

Finally, we run the simulation at a fixed 5 years at half-duty cycle. Varying observation time has a straightforward linear effect on sensitivity, i.e. doubling observation time doubles sensitivity. Therefore, we exclude this parameter from consideration. Additionally, we keep a fixed bandwidth of 18\,Mhz centered on redshift $z=30$ with lower bound $z\sim25$ and upper bound $z\sim38$. This choice of bandwidth is discussed further in the Appendix.

\begin{figure}[ht!]
\plotone{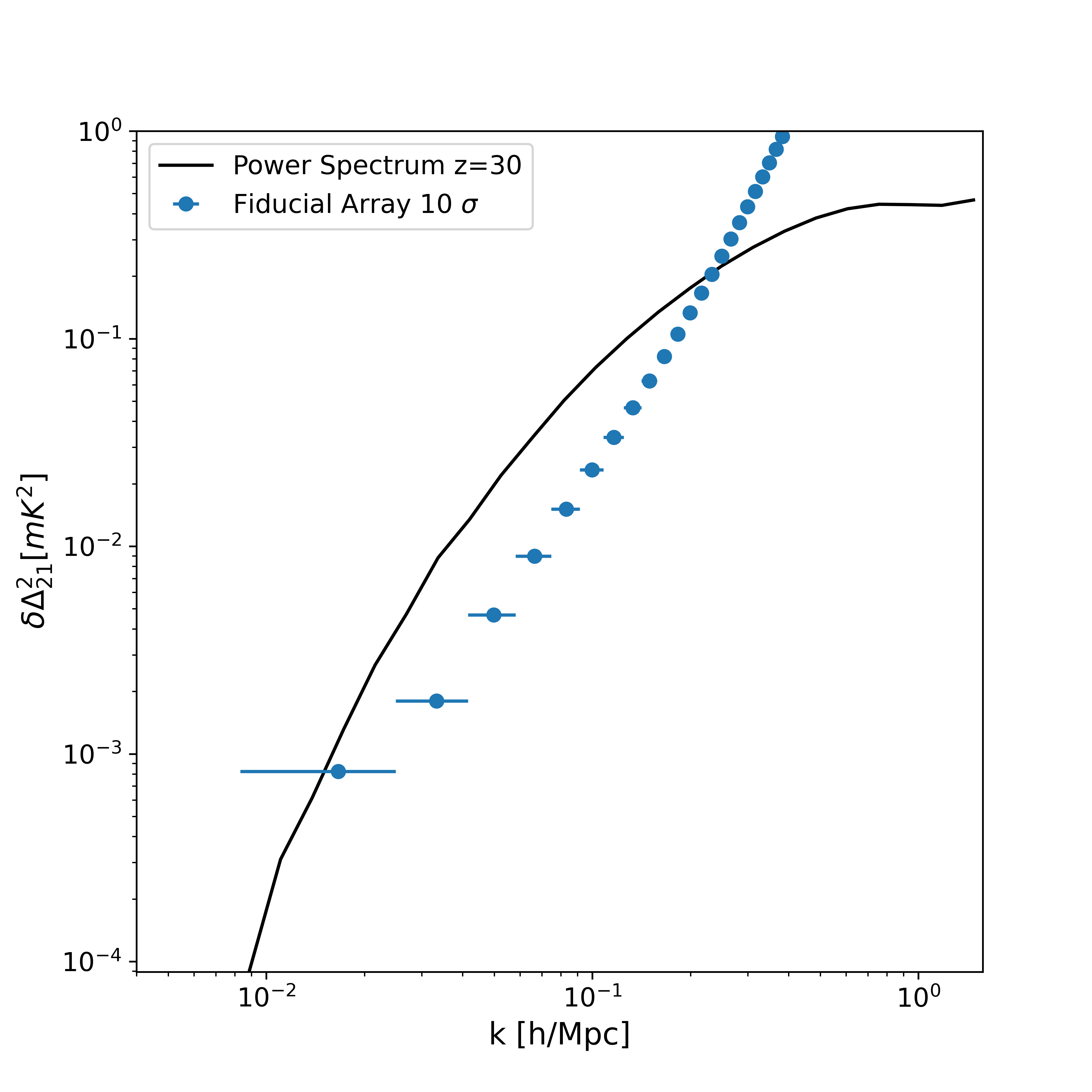}
\caption{One dimensional sensitivity (blue) showing a 10.34$\sigma$ detection against the standard power spectrum (black) at redshift 30. Bins, shown as horizontal lines with a dot representing the center, have a size inversely proportional to the bandwidth. The same format is used for all subsequent sensitivity plots where blue dotted bins represents the fiducial array. The significance of detection of each measurement is given in the legend.} 
\label{fig:fiducial}
\end{figure}

We show the sensitivity of the fiducial array in Figure \ref{fig:fiducial}. The noise $\delta \Delta^{2}_{21}$ in $\rm{mK}^{2}$ at each 1D k-mode in $h/\rm{Mpc}$, represented by a blue dot, is averaged in bins of equal width inversely proportional to the bandwidth. We directly compare this to the 1D dimensionless power spectrum $\Delta^{2}_{21}$ at $z=30$ shown by the solid black curve. For consistency in all figures henceforth, the fiducial array's noise is represented by bins centered by blue dots and the power spectrum is represented by a solid black curve. The fiducial array yields a significance of detection of 10$\sigma$ at $z=30$. Without tracking, the significance of detection is reduced to 3.4$\sigma$. 

\subsection{Subarray Size} \label{subsec:subarray}

Changing the number of dipoles $n$ in a subarray has two effects on sensitivity. The first results from the tight packing of subarrays, since $n$ corresponds directly to a subarray's total side-length $r_{s}$ which determines the center-to-center spacing of subarrays. Therefore, the smallest accessible baseline, also with length $r_{s}$, varies based on $n$. Second, and more straightforwardly, changing $n$ changes  a subarray's $A_{\rm eff}$, limiting or expanding the field of view. 

\begin{figure}
    \plotone{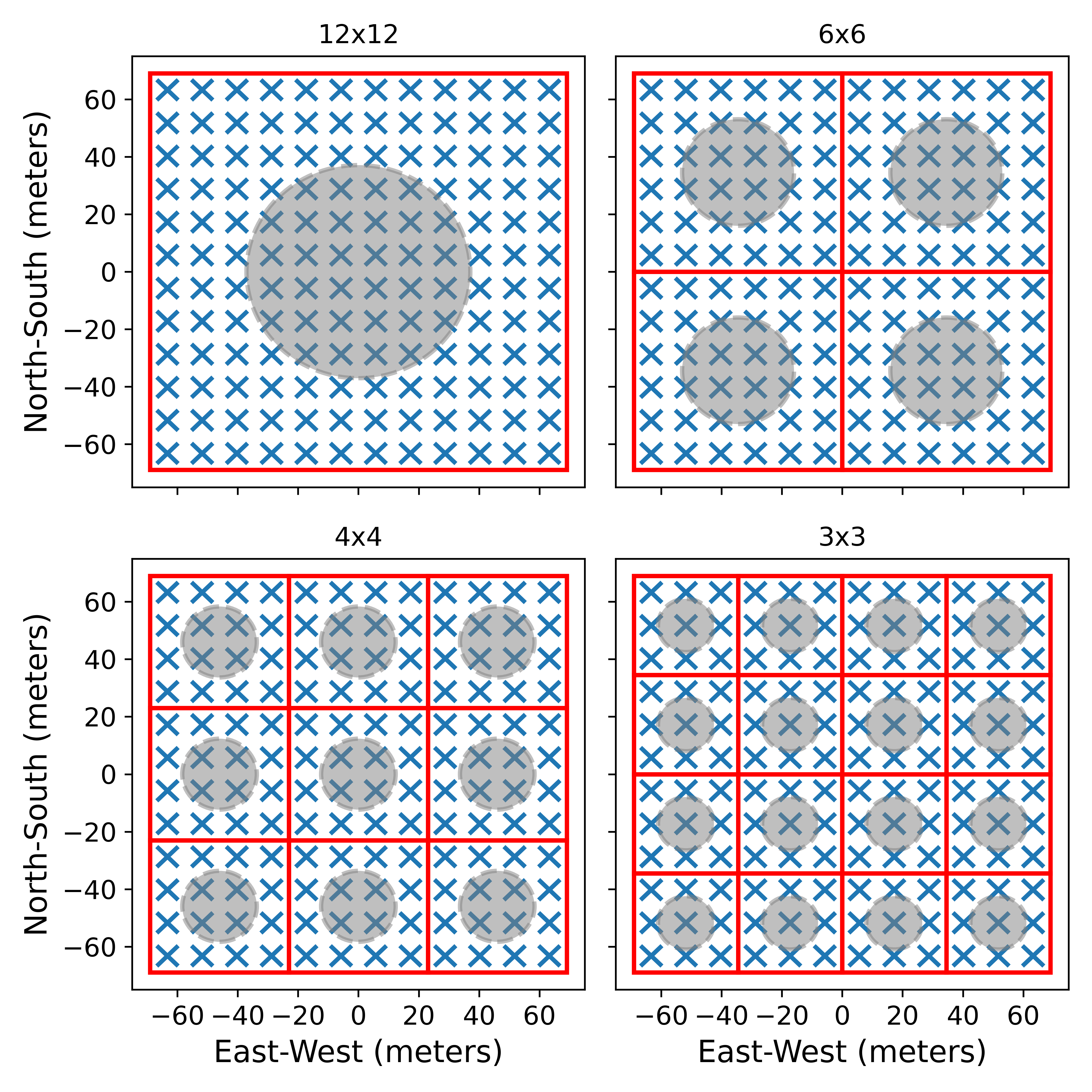}
    \caption{Layouts of a single subarray grouping in a cluster with four different subarray dimensions: 12x12, 6x6, 4x4 (fiducial), and 3x3. The dimensions maintain 144 dipoles per grouping thus allowing clusters to have 576 dipoles in all layouts. \texttt{21cmSense} antennas are shown per subarray with $d_{A}=74.2\,\rm{m},37.1\,\rm{m},24.7\,\rm{m},18.5\,\rm{m}$ for 12x12, 6x6, 4x4, and 3x3 dimensions respectively.}
    \label{fig:subarrayDimensions}
\end{figure}

In this section, we study the effect on sensitivity from different subarray sizes of the fiducial array by varying $n$ --- where we use the notation $\sqrt{n}$ x $\sqrt{n}$ to denote different subarray sizes. We maintain the same number of dipoles in each cluster, composed of four groupings, and thus have the same total 82,944 dipoles for the entire array. However, changing $n$ also changes the number of correlations. Because a 3x3 dipole subarray is on the same order of $N^{2}$ as the fiducial array, we include it in our study but otherwise do not simulate the smaller subarray size of 2x2 as $N^{2}$ begins to exceed $10^{8}$, pushing computational constraints. Therefore, by maintaining the same number of dipoles in a cluster grouping, we are only able to include additional 12x12 and 6x6 dipole subarrays in our study, where a 12x12 dipole subarray is equal to a single grouping in a cluster. These configurations for a single grouping are shown in Figure \ref{fig:subarrayDimensions}. Note that the \texttt{21cmSense} simulated antenna diameter $d_{A}$ changes as a function of subarray dipole count.

\begin{figure}[ht!]
\plotone{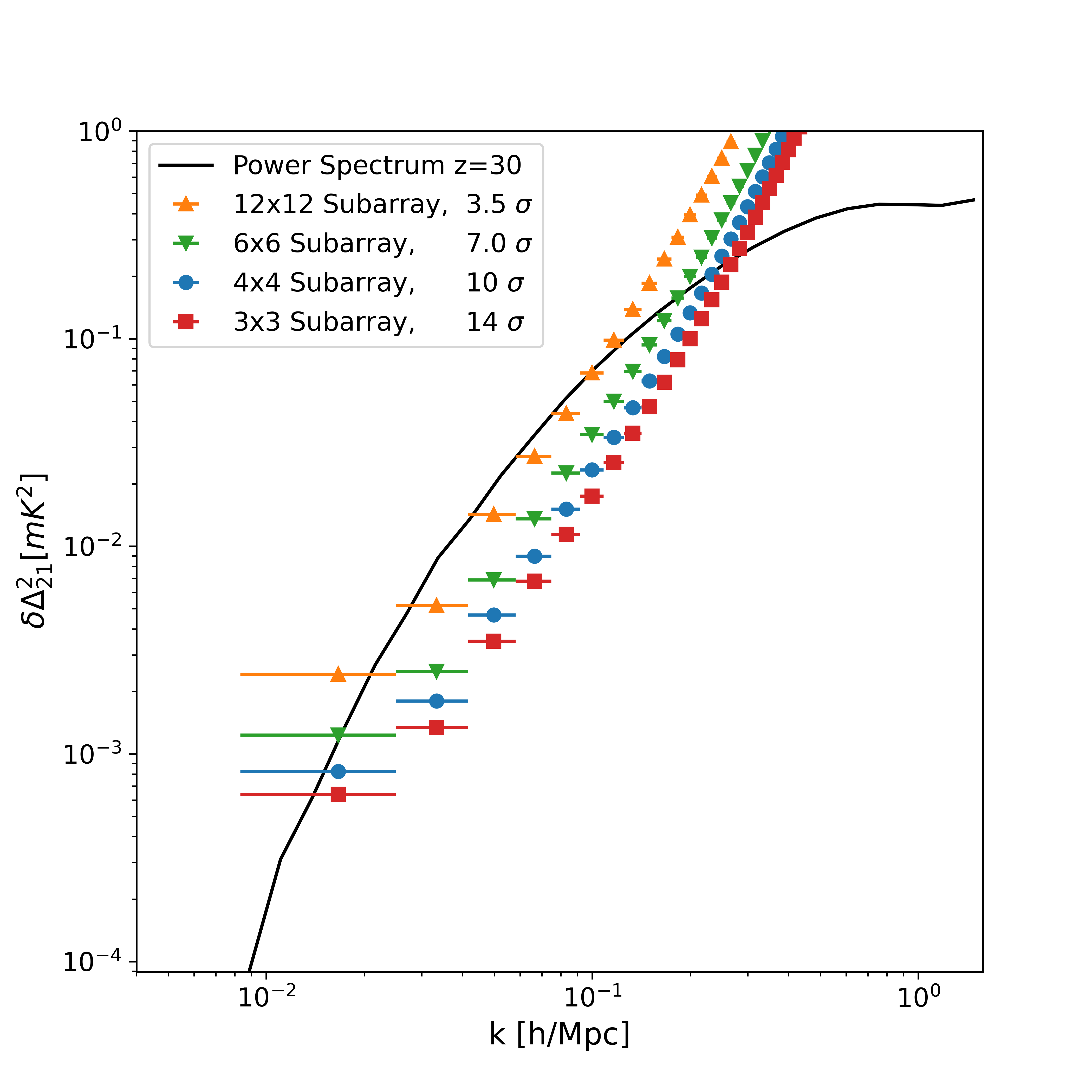}
\caption{Comparison of designs using 3x3, 4x4, 6x6, and 12x12 dipole subarrays as depicted in Figure \ref{fig:subarrayDimensions}. The significance of detection of each measurement is given in the legend. The total number of dipoles in the array remains the same.} 
\label{fig:subarray}
\end{figure}

We then compare the fiducial array to the 3x3, 6x6, and 12x12 dipole subarray configurations in Figure \ref{fig:subarray}. We see a minimum significance of detection of 3.5$\sigma$ and a maximum of 14$\sigma$ for the simulated range.

If we trace how $r_{s}$ evolves, which is directly proportional to the number of dipoles along a side, we find that the significance of detection evolves inversely. In other words, halving the subarray side-length doubles the significance of detection. Notably, the 12x12 and 6x6 dipole subarray configurations offer more modest $N^2 \sim 10^{5}$ and $N^2 \sim 10^{6}$ correlations respectively, so these configurations would not be entirely obsolete in arrays with orders of magnitude more individual dipoles.

\subsection{Dipole Spacing} \label{subsec:spacing}

\begin{figure}
    \plotone{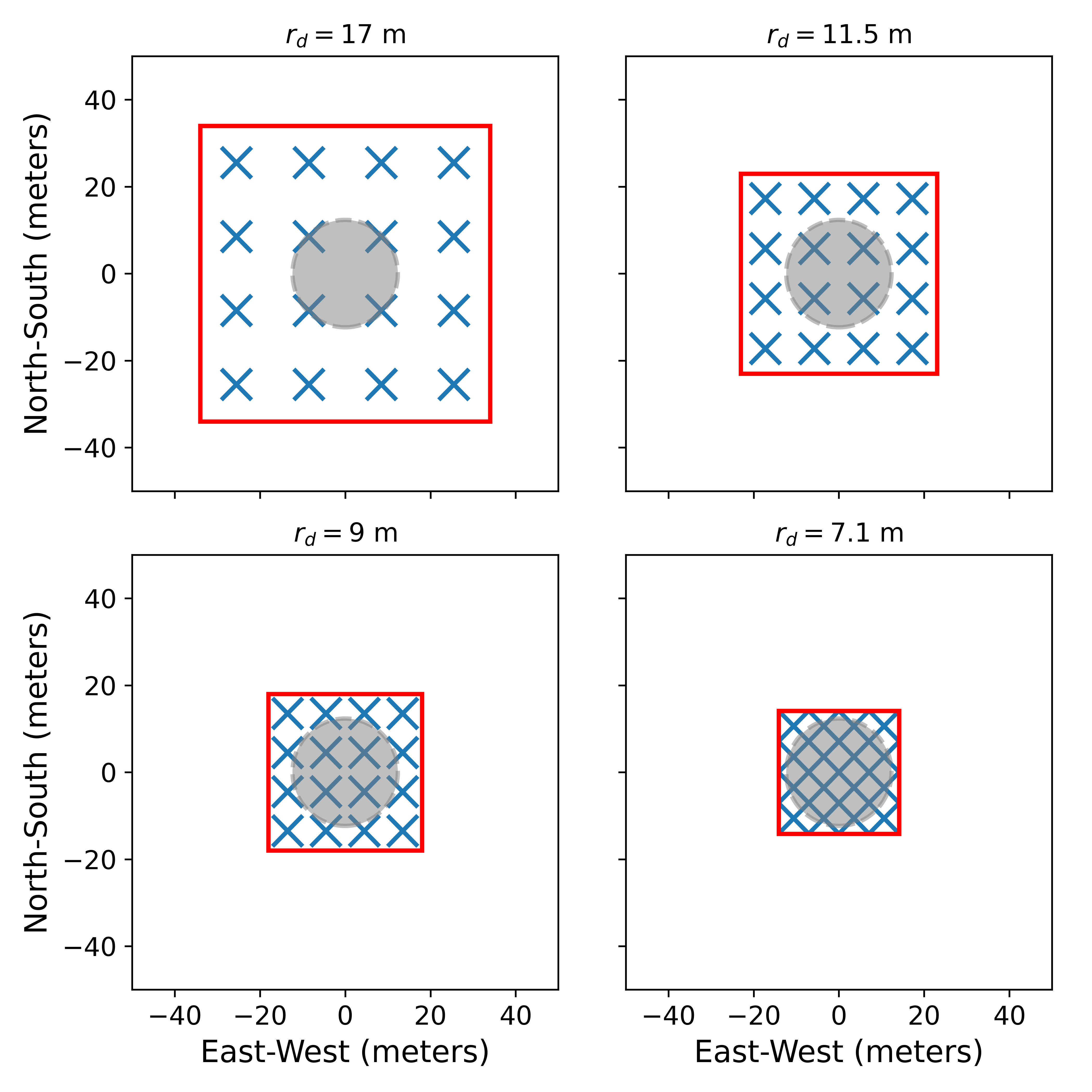}
    \caption{Layouts of a single subarray with four different dipole spacings: $r_{d}=17\,\rm{m},11.5\,\rm{m},9\,\rm{m},7.1\,\rm{m}$. Note that the \texttt{21cmSense} antenna has diameter $d_{A}=24.7\,\rm{m}$ in all subarrays and nearly fills the smallest dipole spacing layout of $r_{d}=7.1\,\rm{m}$. }
    \label{fig:subarraySpacings}
\end{figure}

We also vary the dipole spacing $r_{d}$ to study its effect on the sensitivity. This has one primary effect in changing the center-to-center spacing $r_{s}$, and thus changing the smallest accessible baselines. $A_{\rm eff}$ and $d_{A}$ are unchanged because the dipoles are kept at fixed length and number. However, because of this, we can only pack the dipoles so tightly, i.e. until the tip-to-tip spacing between dipoles is roughly zero. Therefore, the smallest allowable spacing becomes $r_{d}\sim7.1$\,m, the width of the X-shaped dipole. This spacing would prevent a lunar rover from servicing the array, but we include it for a useful lower limit. We also include $r_{d}=17$\,m --- the dipole spacing of the original ASR concept --- and $r_{d}=9$\,m to examine the trend between $r_{d}$ and sensitivity. These layouts are shown in Figure \ref{fig:subarraySpacings}.

\begin{figure}[ht!]
\plotone{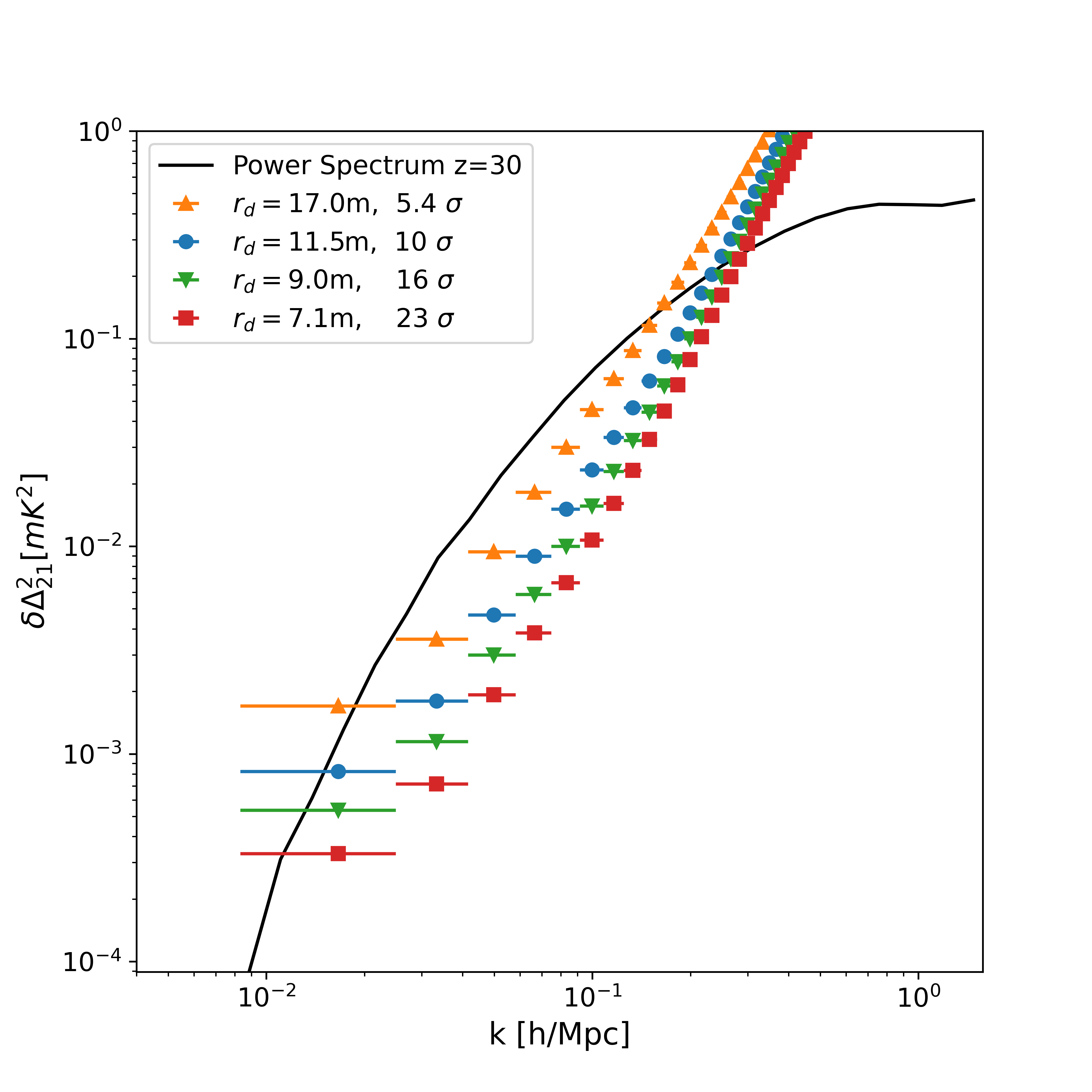}
\caption{A comparison of arrays with different dipole spacing as depicted in Figure \ref{fig:subarraySpacings}. 17\,m spacing is included to compare to the original FarView dipole spacing. 7.1\,m spacing is representative of an array with 10\,m dipoles touching.} 
\label{fig:spacing}
\end{figure}

We show how the sensitivity evolves as a function of $r_{d}$ in Figure \ref{fig:spacing}. We find a minimum significance of detection of 5.4$\sigma$ and a maximum of 23$\sigma$ for the simulated range. While the sensitivity is inversely proportional to $r_{d}$, it is not as straightforward as the dependence on $n$ in the previous section. 

Interestingly, the $r_{s}$ between the 6x6 subarray and $r_{d}=17$\,m subarray and between the 3x3 subarray and $r_{d}=9$\,m subarray are very similar. However, the 6x6 subarray outperforms its counterpart while the 3x3 subarray underperforms its counterpart in sensitivity. Recalling that $r_{s}$ sets the smallest baselines, this shows that for similar baseline distributions, a larger antenna is favored.

\subsection{Collecting Area} \label{subsec:area}

Recall that we treat the total collecting area as the core area only i.e. $A_{\rm coll}= A_{\rm core}$. We simulate change in $A_{\rm coll}$ by adding or subtracting clusters from the edges of the array. We compare grids of 10x10, 14x14, 16x16, and 18x18 clusters to our fiducial 12x12 cluster array depicted in Figure \ref{fig:fiducialArray}. This range corresponds to a minimum collecting area of $A_{\rm coll}=1.73\,\rm{km}^{2}$ and maximum of $A_{\rm coll}=5.60\,\rm{km}^{2}$. The total number of dipoles range from 57,600 dipoles to 186,624 dipoles. Otherwise, all other parameters are fixed.

\begin{figure}[ht!]
\plotone{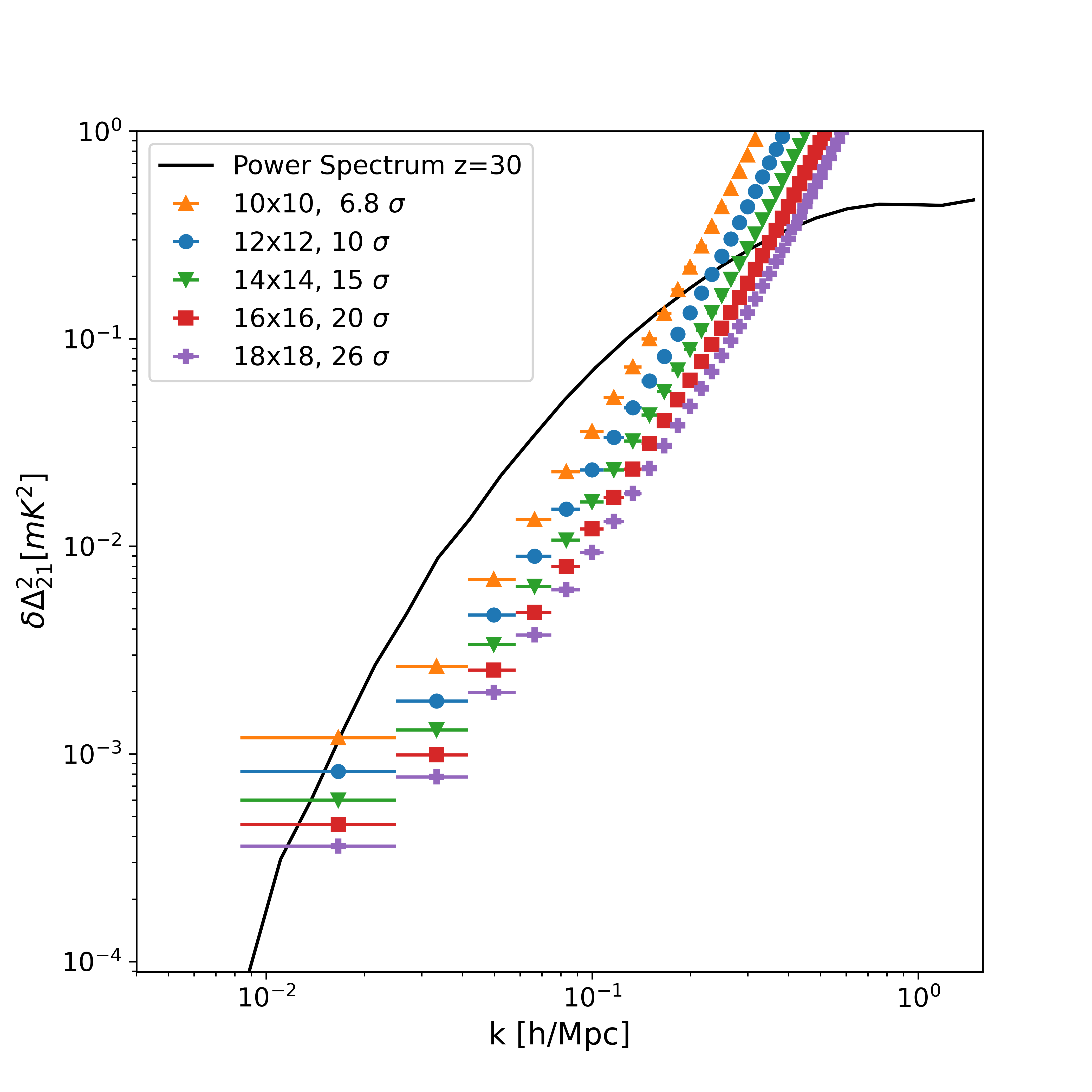}
\caption{A comparison of arrays with different total collecting areas represented by a change in the dimensions of clusters making up the array: 10x10, 12x12, 14x14, 16x16, and 18x18. A linear trend is established between collecting area and sensitivity.} 
\label{fig:collecting_area}
\end{figure}

We compare the sensitivities of these array configurations in Figure \ref{fig:collecting_area}. We see a minimum significance of detection of 6.7$\sigma$ and maximum of 26$\sigma$ for the simulated range. Of note, the 14x14 grid layout contains 112,896 dipoles which sits comfortably around the number of dipoles the FarView engineering teams expect to be able to deploy and offers a 4.33$\sigma$ increase in significance of detection over the fiducial array.

From Equations \ref{eq:Noise} and \ref{eq:Trms}, we expect an inverse dependence between noise and total collecting area, or a linear dependence between significance of detection and collecting area. As anticipated, we find this trend in Figure \ref{fig:collecting_area} with a $\sim \rm 0.08\sigma$ increase in significance of detection per additional cluster. This is useful because we are limited in the smallest correlated element size and in how tightly we can pack the array, but we can always add to the total collecting area much like we can always increase the total observing time. 

\subsection{Foregrounds} \label{subsec:fore}

So far we have considered sensitivities without any contamination from foregrounds. Realistically it is difficult to entirely remove foregrounds from the signal. In order to understand this, consider the 2D Fourier space with $k_{\parallel}$ representing line-of-sight modes and $k_{\perp}$ representing transverse sky modes. In principle, one might expect spectrally smooth foregrounds to occupy only the lowest $k_{\parallel}$ bins. However, because the angular Fourier mode sampled by a baseline changes with frequency, $k_{\parallel}$ and $k_{\perp}$ are intrinsically coupled. This leads to the leakage of spectrally smooth foregrounds from low $k_{\parallel}$ into a wedge-shaped region in 2D $k$-space that increases in $k_{\parallel}$ with increasing $k_{\perp}$ according to a power law. For a more in-depth review of the foreground wedge refer to \citet{Liu_2020}.

\texttt{21cmSense} offers several types of foreground models, represented as different wedge cuts --- i.e., each model excludes a different region of 2D $k$-space from the sensitivity calculation to account for modes irreparably contaminated by foreground emission. We consider two models, one for foreground subtraction (where some modes within the wedge are recovered) and one for foreground avoidance (where only modes outside the wedge are used in the sensitivity calculation).  In our foreground subtraction model (corresponding to the ``optimistic'' model in \texttt{21cmSense}), only modes below the full-width half-max of the primary beam in 2D $k$-space are used (see \citealt{pober_2016} for a discussion of how the primary beam fits into the wedge paradigm); in our foreground avoidance model (corresponding to the ``moderate'' model in \texttt{21cmSense}), the foreground cut extends 0.1 $h \rm{Mpc}^{-1}$ beyond the maximum geometric extent of the wedge (sometime called the horizon limit) \citep{Pober_2014}. 

\begin{figure}[ht!]
\plotone{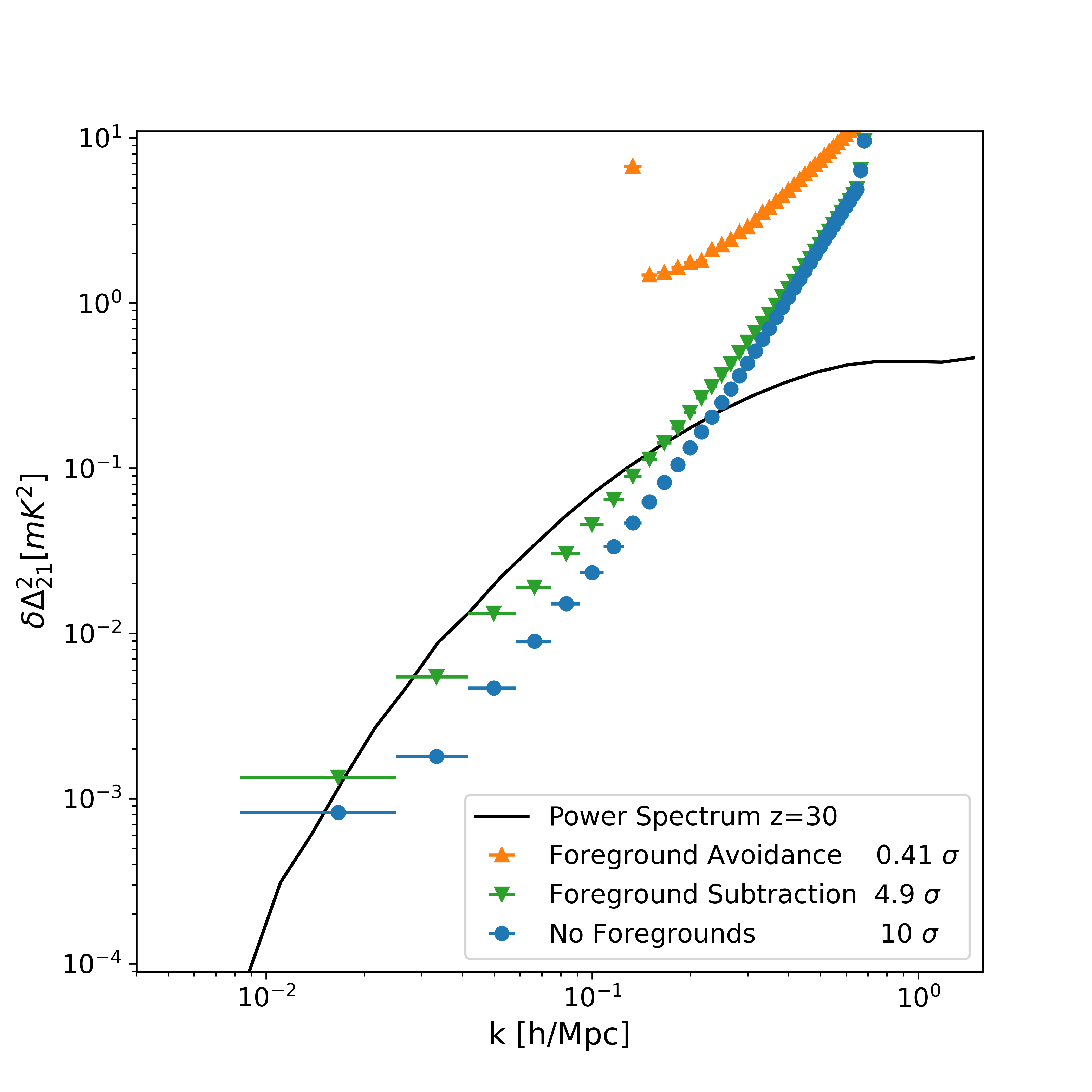}
\caption{A comparison of the fiducial array with and without foregrounds. The $0.1\,h{\rm Mpc}^{-1}$ cutoff in a foreground avoidance strategy removes a significant number of small k modes.} 
\label{fig:Foregrounds}
\end{figure}

We apply these foreground models to the sensitivity calculations with our fiducial array. In Figure \ref{fig:Foregrounds} we see a 4.9$\sigma$ detection with foreground subtraction and a 0.41$\sigma$ detection with foreground avoidance. The subtraction model leaves many low $k$ modes intact, so applying any of the previous methods we've described of increasing small baselines would be able to recover a substantial detection. 

On the other hand, the avoidance model has a very significant effect on sensitivity, effectively precluding a detection without an array substantially larger than the fiducial array.  Furthermore, compared with Epoch of Reionization and other lower redshift 21\,cm experiments \citep{Pober_2014}, the avoidance model causes a larger degradation of the sensitivity.  This is largely the result of the evolution of the wedge with redshift, as discussed in \citet{10.1093/mnras/stu2575}; the impact on redshifts above $z=30$ is expected to be even larger, but we defer a detailed exploration of this effect to future work.

\section{Discussion and Conclusion} \label{sec:conc}

We find that a large-scale radio interferometer to detect the $21\,cm$ power spectrum of the Dark Ages is an ambitious task, but, at least in terms of sensitivity, is ultimately feasible. Our fiducial array which offers a $>10\sigma$ detection at z=30 with only a 5-year lifetime and 2.5 ${\rm km}^2$ of collecting area can be improved upon in numerous ways. 

We find that shrinking the spacing $r_{d}$ between dipoles and decreasing the number of dipoles $n$ in a subarray, or single correlated element, both offer the greatest increases in sensitivity. This increase comes from the addition of small baselines that substantially decrease the noise at large scales. Notably, in the case of similar baseline distribution, a subarray with more dipoles, and thus larger $A_{\rm eff}$, is more sensitive than a subarray with more tightly packed but fewer dipoles despite the former offering a smaller FoV. However, it is also useful to note that we ignore the electromagnetic coupling between tightly spaced dipoles in preforming our sensitivity calculations which may greatly impact an instrument's ability to detect the 21\,cm signal (see e.g. \citealt{rath_et_al_2024}). Thus, the trade-off between $r_{d}$ and $n$ is something that requires more study in complex simulations.

Of course, $r_{d}$ and $n$ come with limits. We can only push dipoles together so much before they are overlapping, and we face a more significant computational workload the fewer dipoles we pack into one subarray. This pushes us to consider the more traditional methods of increasing sensitivity: increasing collecting area $A_{\rm coll}$ and observation time.

We find that increasing $A_{\rm coll}$ for a lunar array behaves as expected for a typical radio interferometer. It is useful to note that the addition of 100,000 more dipoles to our fiducial array increases the significance of detection by $\sim 15\sigma$. Thus, doubling the array size, slightly more than doubles the sensitivity. It is of course, a larger task to manufacture twice as many dipoles than to pack subarrays more tightly. So, we find it is more useful to have an initial tightly packed array before adding additional dipoles. 

Notably, the noise at small scales is barely diminished by varying our design parameters. Thus, while we find that a detection of the 21\,cm Dark Ages signal is possible with current array concepts, further work is necessary to acquire the sensitivity to observe small-scale structure in the $21\,\rm{cm}$ power spectrum of the Dark Ages. However, exotic physics such as charged dark matter has substantial increases on large-scale power which would likely be observable at z=30 given our fiducial array \citep{Mu_oz_2018}. We believe it is therefore possible to probe exotic physics with an array of this scale, but a full study of parameter constraints is beyond the scope of the current work.

In regard to prior sensitivity estimates of Lunar far side arrays, we find stark disagreement with the ASR FarView concept's sensitivity prediction. While it is true that the ASR concept sensitivity is based on an approximation which assumes a constant density of visibilities in the $uv$-plane, this reason alone is not sufficient to explain such a large difference. Implicitly, the analytic approximation assumes an all-sky field of view, which requires every dipole be correlated, such as in the CoDEX array \citep{koopmans2019peeringdarkageslowfrequency}. While computational constraints prevent us from simulating such an array, we can extrapolate the trend from Section \ref{subsec:subarray} down to an array where every dipole is correlated, or $n=1$ per subarray. After adjusting for the difference in $A_{\rm{coll}}$, we find that this extrapolation is on the same order of sensitivity forecasts for the $10\,\rm{km}^{2}$ CoDEX model i.e. $\delta\Delta^{2}_{21}\sim10^{-3}\,\rm{mK}$ at $k=10^{-1}\,\rm{h/Mpc}$. Therefore, we do find agreement in the sensitivity between \texttt{21cmSense} and the analytical approximation used for FarView and CoDEX in the case of a tightly packed array.

Lastly, we again note that achieving our predicted sensitivities requires foreground \emph{subtraction} (c.f. Figure \ref{fig:Foregrounds}).  The foreground avoidance approach used by many 21\,cm EoR experiments is not suitable for the Dark Ages, an effect we will explore in a future work.  Here we simply point out that, while foreground subtraction is daunting, the challenge is not necessarily the same as it is on the Earth.  Ionospheric distortions of the foreground emission require time-dependent (and possibly antenna-dependent) updates to the foreground model, whereas on the lunar far side, a single high-precision foreground model may suffice to describe all observations.  Further research is required to determine whether a sufficiently accurate foreground catalog could be constructed under these conditions.

\begin{acknowledgments}

The authors acknowledge support for this work from NASA grants 80NSSC18K0389, 80NSSC21K0693, and 80NSSC22K1745.  WS also acknowledges support from a NASA Rhode Island Space Grant Fellowship.    We would also like to thank Jack Burns and Ron Polidan for technical assistance with the FarView design and Adrian Liu and Frank Ning for helpful conversations that aided in the preparation of this work. 
\end{acknowledgments}

\vspace{5mm}

\software{21cmSense \citep{Murray2024}, 21cmFast \citep{10.1111/j.1365-2966.2010.17731.x, Murray2020}, astropy \citep{astropy:2013, astropy:2018, astropy:2022}, lunarsky (\url{ https://github.com/aelanman/lunarsky}), pyuvdata \citep{Hazelton2017}}

\appendix

The question of what bandwidth is appropriate for 21\,cm cosmology is a subtle one.  The total bandwidth used in an observation corresponds to the range of redshifts that are observed in the 21\,cm line.  Once the line-of-sight Fourier transform over frequency is performed, however, all redshifts become mixed together --- that is, every $k_\parallel$ mode contains information from all the redshifts.  If too large a bandwidth (i.e too large a redshift range) is used, then the statistics of the 21\,cm signal will not be constant across the band.  This can complicate comparison with theoretical models, where simulations of the 21\,cm field at a single redshift (``co-eval boxes'') are often used.  Therefore, most analyses (and most sensitivity calculations) focus on a narrow redshift range so that evolution across the band can be neglected.  Typical values for EoR experiments are 6\,MHz \citep{Parsons_2012, McQuinn_2006} or $\Delta z = 0.5$ \citep{Pober_2014}.  Multiple sub-bands centered on different redshifts are then used to constrain the evolution of the power spectrum.

At Dark Ages redshifts, however, such a rule of thumb becomes impractical.  At $z=30$, a $\Delta z$ of 0.5 corresponds to $\sim0.75$\,Mhz.  Enforcing measurements to come from such a small bandwidth drastically limits the smallest $k_\parallel$ mode that can be measured (since small $k_\parallel$ values correspond to slowly evolving modes along the line-of-sight).  Furthermore, this restriction is, ultimately, artificial.  There \emph{is} information in these large-scale $k_\parallel$ modes, only extracting it can be complicated by evolution across the redshift band.  This evolution, however, can be forward modeled; for example, the latest version of the \texttt{21CMMC} code from \citet{greig_and_mesinger_2018} simulates evolving lightcones spanning large redshift ranges (rather than coeval boxes) to compare with power spectra made from a wide bandwidth.  This forward-modeling is, in principle, even more straightforward for the Dark Ages, where the signal and its evolution can be analytically calculated using a Boltzmann code.  

For the purposes of the present work, we assume that such forward modeling will be possible, enabling us to recover scientific information from large-scale line-of-sight modes.  Our sensitivity calculations use an 18\,MHz bandwidth centered at $z=30$ (45.8\,MHz), chosen to span a wide range but to include no information from $z < 25$ where, in our \texttt{21cmFAST} simulations, astrophysical effects start to dominate the 21\,cm power spectrum.  For simplicity, when calculating both the fiducial power spectrum and thermal noise level, we use only the power spectrum and sky temperature from the center redshift $z=30$.  In practice, the average power spectrum over the band is 45\% higher than this value, while the average noise is only 13.7\% higher.  Ultimately, then, this assumption is conservative from a sensitivity perspective, and somewhat higher significances may be attainable.

\bibliography{references}{}
\bibliographystyle{aasjournal}

\end{document}